# HD 240121 - an ACV variable showing anti-phase variations of the $B$ and $V$ light curves

Rainer Gröbel[a], Stefan Hümmerich[a,b], Ernst Paunzen[c], Klaus Bernhard[a,b]

[a]*Bundesdeutsche Arbeitsgemeinschaft für Veränderliche Sterne e.V. (BAV), Berlin, Germany*
[b]*American Association of Variable Star Observers (AAVSO), Cambridge, USA*
[c]*Department of Theoretical Physics and Astrophysics, Masaryk University, Kotlářská 2, 611 37 Brno, Czech Republic*

**Abstract**

The variability of HD 240121 = BD+59 2602 was first suspected by Särg & Wramdemark (1970) and later confirmed by Gröbel (1992a,b). Because of the observed anti-phase variations of the $B$ and $V$ light curves, the latter author tentatively suggested an ACV type. Apart from its inclusion in the catalog of New Suspected Variables (NSV 25977), no further investigations of the star have been published. HD 240121 was included into our target list of ACV candidates and investigated in order to determine the reason for the observed brightness variations. All available information on HD 240121 were collected via an exhaustive data mining procedure. Data from Gröbel (1992a,b) were re-analysed and photometric observations from the NSVS and Hipparcos archives were procured and investigated. Line-of-sight reddening and stellar parameters were calculated from archival photometric data. HD 240121 is a young, late B-type CP2 star of the silicon subgroup. The observed period, amplitude of light variations and variability pattern (anti-phase variations) are typical of ACV variables. The occurrence of anti-phase variations of the $B$ and $V$ light curves is rarely observed and points to the existence of a null wavelength in the visual spectrum. We therefore strongly encourage further multi-colour photometric observations of this star.

## 1. Introduction

Magnetic, chemically peculiar (CP2) stars exhibit a nonuniform distribution of chemical elements, which is believed to be produced by selective processes (radiative levitation, gravitational settling) operating in calm radiative atmospheres (Richer et al. 2000). In these stars, the formation of spots and patches of enhanced element abundance are observed (Michaud et al. 1981), which is likely connected with the influence of a strong magnetic field.

Many CP2 stars exhibit strictly periodic changes in their spectra and brightness in different photometric passbands, which are satisfactorily described by the oblique rotator model (Stibbs 1950). Photometrically variable CP2 stars are referred to as $\alpha^2$ Canum Venaticorum (ACV) variables in the GCVS (Samus et al. 2007 – 2015). The observed periodicity of variation is the rotational period of the star. The rotational periods of ACV variables tend to cluster around a maximum value of $P \approx 2$ days (Renson & Manfroid 2009; Bernhard et al. 2015a).

The observed photometric variability is thought to result from a redistribution of flux in the surface abundance spots (e.g. Krtička et al. 2013). Amplitudes of several hundredths up to some tenths of magnitude are observed (Mathys & Manfroid 1985), and the amplitude of the light variability generally increases towards shorter wavelengths.

Furthermore, the observed amplitude and shape of the light curve are highly dependent on the investigated spectral region. While the light changes may be in phase in different photometric passbands, the flux might also remain almost unchanged at certain wavelengths (null wavelength; Molnar 1975) or vary in anti-phase to the flux at other wavelengths (e.g. Manfroid & Mathys 1986; Shulyak et al. 2010). ACV variables show an amazing diversity of light curves, depending on the distribution of photometric spots on the surface of the star and the mechanism of variability which is connected to the elements involved (e.g. Mikulášek et al. 2007).

## 2. Target Star – Historical Data and Observations

The variability of HD 240121 = BD+59 2602 = TYC 4278-969-1 = NSV 25977 (RA, Dec (J2000) = 22h 55m 50s.344, 60° 00' 34".53; Tycho-2 position) was first suspected by Särg & Wramdemark (1970) during their $UBV$ photometric survey of early type stars in a Milky Way field in Cepheus. From five observations taken between August and October 1969, the authors deduce $V = 9.56$ mag and $(U - B) = -0.01$ mag. The $(B - V)$ index, however, was found to vary between $+0.14$ and $+0.28$ mag.

It is noteworthy to point out that in the aforementioned investigation, the object (listed with the running number 104 by the authors) was misidentified as BD+59 2604. However, the identification with BD+59 2602 is secure from the chart in Särg & Wramdemark (1970), as has been pointed out by the GCVS team (Samus et al. 2007 – 2015).

The findings of Särg & Wramdemark (1970) were followed up by one of us (R. Gröbel). During several nights in 1991 and 1992, HD 240121 was monitored in Johnson $B$ and $V$ using a 931A photomultiplier on an 8-inch reflecting telescope. 122 measurements in $V$ and 121 measurements in $B$ were obtained. Details about the instrumentation and reduction processes can be found in Gröbel (1992a,b).

From these observations, light variability of HD 240121 with amplitudes of $\Delta B = 0.08$ mag and $\Delta V = 0.04$ mag was established. A best fit to the observations was achieved with a period of $P = 2.025$ d. Because of the observed anti-phase variations between the $B$ and $V$ light curves, Gröbel (1992b) tentatively suggested that HD 240121 might be an ACV variable and called for spectroscopic observations. On the basis of these results, the star was included in the catalog of New Suspected Variables as NSV 25977 (Samus et al. 2007 – 2015). No further observations of the star have been published.

## 3. Archival Data, Analysis and Discussion

Interest in NSV 25977 was renewed during the ongoing search for new ACV variables in public sky survey data (Bernhard et al. 2015a,b). The object was investigated using the online catalogue systems SIMBAD (Wenger et al. 2000) and VizieR (Ochsenbein et al. 2000) and was found to be a spectroscopically confirmed CP2 star. Originally classified with a spectral type of B8 in the Henry Draper Catalogue (Cannon & Pickering 1928), it was later identified as a chemical peculiar star of the silicon subgroup by Bidelman (1966). The Catalogue of Ap, HgMn, and Am stars (Renson & Manfroid 2009), which includes HD 240121 under the identification number 59870, gives a spectral type of B9pSi.

### 3.1. Groebel (1992a,b) data

Light curves in $B$ and $V$ and the resulting $(B - V)$ colour curve of the data obtained by Gröbel (1992a,b) are illustrated in Figure 1. The data were reinvestigated using PERIOD04 (Lenz & Breger 2005). The resulting Fourier spectrum (shown in Figure 2) is dominated by peaks at the main frequency $f = 0.4943$ c/d ($P = 2.0232$ d) and $(1-f)$. We are confident that the signal at $f$ represents the true period, as it boasts the highest amplitude and produces a significantly better fit to the data. Furthermore, an analysis of NSVS data (Woźniak et al. 2004) confirms these findings (see below). The following elements were derived; the corresponding folded light curves are shown in Figure 3.

$$HJD(MaxB) = 2448492.925 + 2.0232\ (\pm\ 0.0002) \times E \quad (1)$$

It is important to point out that, because of the small number of observations and the period value close to two days, the period solution has to be viewed as preliminary.

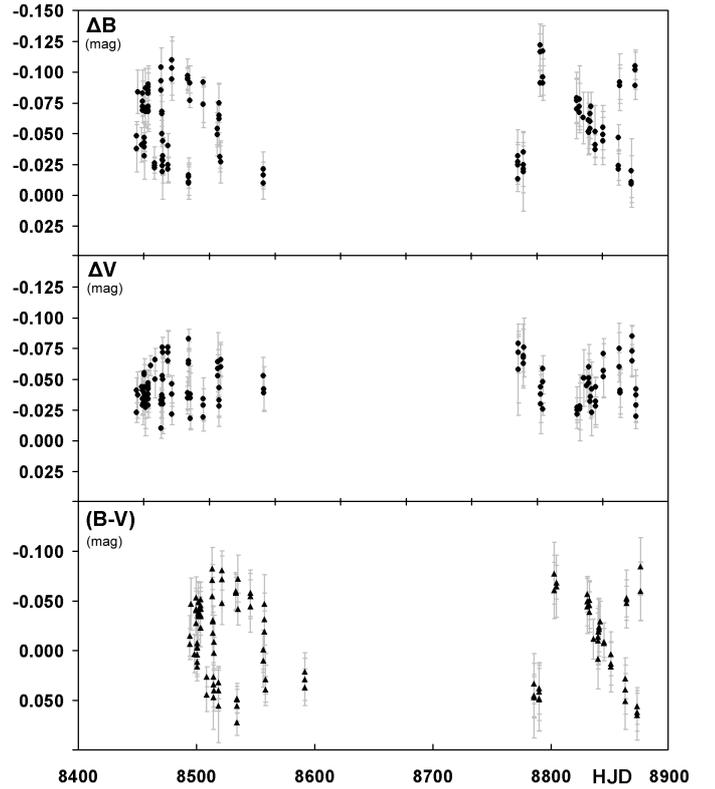

Figure 1: $B$ and $V$ light curves and $(B - V)$ colour curve of HD 240121, based on data obtained by Gröbel (1992a,b). Time is given in HJD-2440000.

It cannot be excluded that the derived value represents an alias of the true period. This, however, is unlikely, as the period is confirmed in NSVS data.

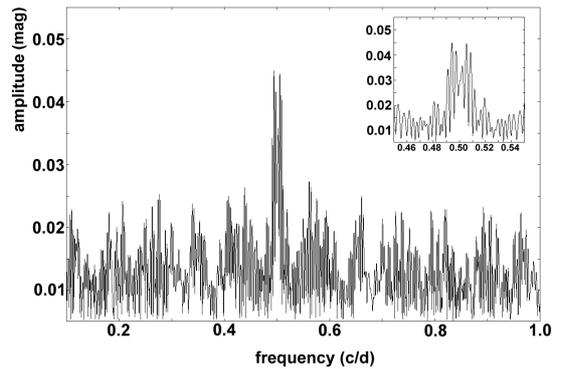

Figure 2: Fourier spectrum of Gröbel (1992a,b) $B$ data for HD 240121 in the investigated frequency range of $0.1 < f(c/d) < 1$. The inset shows a detailed view of the frequency region around 0.50 c/d.

Figure 3 illustrates that the brightness of HD 240121 in $B$ and $V$ varies approximately in anti-phase, with the maximum in $B$ occurring at a phase of $\varphi \approx 0.0$ and the maximum in $V$ at a phase of $\varphi \approx 0.5$. This leads to the observed unusual variations in the $(B-V)$ index. The observed anti-phase variations exclude pulsational variability as source of the observed light changes and are strong evidence for the star's classification as an ACV variable. This is further corroborated by the proposed period ($P = 2.0232$ d) and the observed amplitudes ($\Delta B = 0.08$ mag and $\Delta V = 0.04$ mag; $\Delta B > \Delta V$), which are typical of ACV variables.

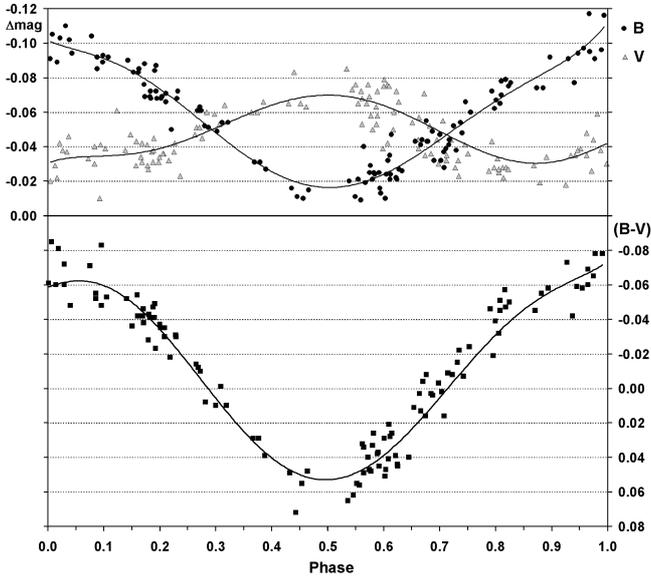

Figure 3: Folded light curves in $B$ and $V$ and the folded $(B-V)$ colour curve of HD 240121, based on the elements given in Equation (1). The solid lines indicate a 6th order polynomial fit to the data.

While anti-phase variations between ultraviolet and visual light curves of CP2 stars are regularly documented (e.g. Molnar 1973), this phenomenon is rarely observed in the visual range and implies the existence of a region in the visual spectrum in which the flux does not vary appreciably over the rotational period (e.g. Kurtz et al. 1996), probably between the $B$ and $V$ passbands. The observed variability of HD 240121 is therefore remarkable and should be a challenge to theoreticians to model.

### 3.2. NSVS and Hipparcos data

In order to further investigate this matter, photometric observations from the NSVS and Hipparcos (van Leeuwen et al. 1997) archives were procured. NSVS data were collected with the ROTSE-I instruments and comprise unfiltered CCD observations which have been placed onto a $V$-equivalent scale (Akerlof et al. 2000); the resulting passband is approximately $4000 - 9000$ Å. Hipparcos data were collected through the non-standard Hipparcos passband ($Hp$, $\lambda eff = 5170$ Å, $\Delta\lambda = 2300$ Å; Bessell 2005). NSVS and Hipparcos data were cleaned of obvious outliers by visual inspection and searched for periodic signals in the frequency range from $0.1 < f(c/d) < 1$ using PERIOD04.

Although Hipparcos data should, in principle, be well suited to investigate the low-amplitude variability presented by HD 240121 (e.g. Paunzen & Maitzen 1998), no significant signal is present in the investigated frequency range. Weak signals at $f = 0.2115$ c/d ($P = 4.7281$ d) and $f = 0.957775$ c/d ($P = 1.04409$ d) are obviously spurious detections with low signal-to-noise ratios that yield only scatter diagrams when used to phase the data. The same holds true when Hipparcos data are folded with the elements given in Equation (1). The corresponding Fourier spectrum is given in Figure 4.

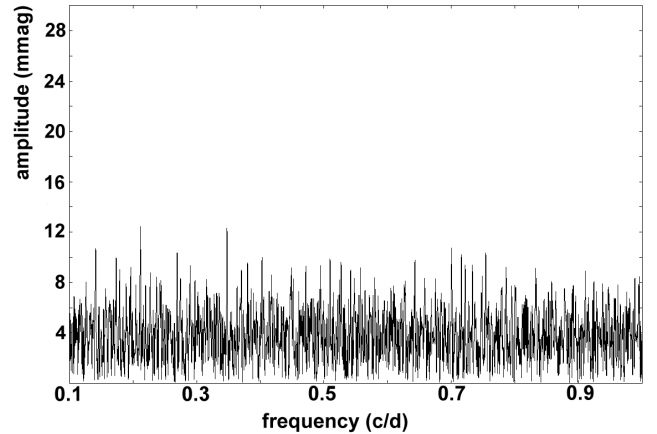

Figure 4: Fourier spectrum of Hipparcos data for HD 240121 = HIP 113233 in the investigated frequency range of $0.1 < f(c/d) < 1$.

As the Hipparcos passband overlaps the standard $B$ and $V$ passbands, it is possible that, because of the observed anti-correlation, the light variations have been smoothed out and the resulting amplitude is too low for detection in Hipparcos data. Furthermore, this might be a first hint at the existence of the proposed null wavelength. Reduced amplitudes of CP2 stars in the Hipparcos passband are e.g. also illustrated in Mikulášek et al. (2007).

The situation is different in NSVS data, which clearly confirm the light variations of HD 240121. The resulting Fourier spectrum is similar to the spectrum calculated from the Gröbel (1992a,b) dataset (cf. Figure 2) and is dominated by peaks at the main frequency of $f = 0.49420$ c/d ($P = 2.02347$ d) and $(1 - f)$. Although these signals are of small amplitude, we are inclined to interpret them as real features because of the very good agreement with the findings from the Gröbel (1992a,b) data. Furthermore, the best fit period agrees well with the period derived from the Gröbel (1992a,b) dataset, thereby confirming our period solution. Fourier spectrum and phase plot based on NSVS data are shown in Figure 5 and 6, respectively.

It is interesting that the light variations of HD 240121 have obviously been detected in NSVS data, which - concerning measurement accuracy - are generally inferior to Hipparcos data for the investigation of small amplitude variability. This might be explained by the different wavelength coverage between both surveys, with the ROTSE-I passband extending further into the red where the variations of the star might be more pronounced. Clearly, an investigation of the light variability of HD 240121 in a multicolour photometric system employing narrower passbands (like e.g. the Strömgren system) would be desirable.

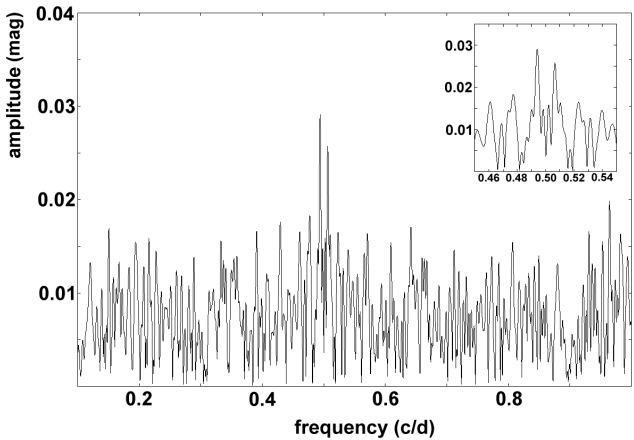

Figure 5: Fourier spectrum of NSVS data for HD 240121 = NSVS 1452417 = NSVS 3469822 in the investigated frequency range of $0.1 < f(c/d) < 1$. The inset shows a detailed view of the frequency region around 0.50 c/d.

### 3.3. Estimation of reddening and stellar parameters

For B and early A-type stars in the solar neighbourhood, the reddening is generally estimated using photometric calibrations such as the $Q$ parameter within the Johnson $UBV$ system (Johnson 1958). This method is only based on photometric indices and does not take into account distance estimates via parallax measurements. The procedure of the $Q$ method is straightforward and has been described in detail by Johnson (1958). The basic correlations are:

$$Q = (U-B) - 0.72(B-V) - 0.05(B-V)^2 \qquad (2)$$

$$E(B-V) = (B-V) - 0.331Q + 0.019 \qquad (3)$$

Assuming $(B-V) = 0.16$ mag and $(U-B) = 0.01$ mag (cf. Section 2), we derive $Q = -0.106$ and $E(B-V) = 0.213$ mag.

To check if this estimate is reasonable, the intrinsic colors of main sequence stars (Pecaut & Mamajek 2013, Table 5) were employed. For the absorption (reddening) in the different passbands, we used the coefficients listed by Fitzpatrick (1999): $A_V = 3.1 E(B-V) = 1.36 E(V-J) = 1.21 E(V-H) = 1.12 E(V-K_s)$. The 2MASS $JHK_s$ photometry for HD 240121 was taken from Skrutskie et al. (2006).

Within a few thousandths of magnitude, the intrinsic colours are consistent with a B9V to B9.5V star which has an effective temperature of about 10 500 K. To verify this result, the calibrations of the effective temperature in terms of $Q$ and $(B-V)$ by Paunzen et al. (2005) and Netopil et al. (2008) were applied. This procedure resulted in 10 200(300) K, which is in line with the former value.

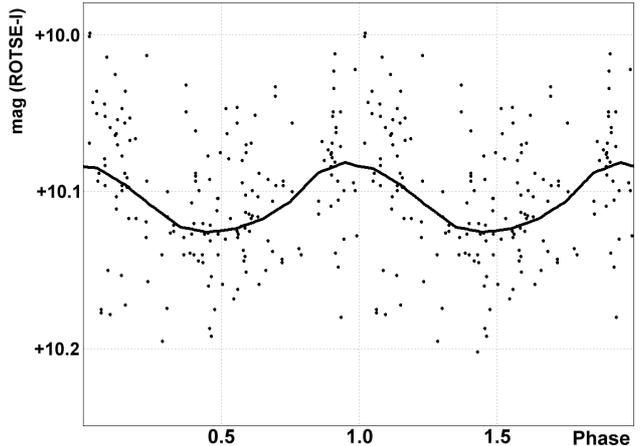

Figure 6: Phase plot of HD 240121 = NSVS 1452417 = NSVS 3469822, based on NSVS data and folded with $P = 2.02347$ d. The solid line indicates the best fit to the data.

As last test, we applied the temperature calibration using $(V - K_s)_0$, which was especially derived for CP stars (Herdin et al. 2016), and yielded 10900(500) K. Again, this is in line with all other values, which provides confidence in this astrophysical parameter.

We want to stress that McDonald et al. (2012) listed an effective temperature of 7 800 K for HD 240121. These authors used a Spectral Energy Distribution (SED) fitting technique which is very sensitive to reddening and the overall stellar flux distribution. However, the latter is very peculiar for CP2 stars which exhibit several flux depressions and a significant UV excess (Stigler et al. 2014). We applied the VO Sed Analyzer (VOSA) in version 5.0 (Bayo et al. 2008) and derived a best fit of 8 000 K with a zero reddening. Using the above derived reddening estimate and an enhanced metallicity, an effective temperature of 10 000 K was derived. This illustrates a possible pitfall when using SED-fitting in an automatic way for many stars as done by McDonald et al. (2012).

The parallax from the Hipparcos measurement for HD 240121 is $\pi = 2.66(95)$ mas which gives a distance between 277 and 585 pc, respectively. The star is located in the Galactic plane at $[l,b] = [108°.953, +0°.313]$. The derived, relatively high reddening for such a nearby object is therefore not surprising. We used $E(B-V) = 0.213$ mag and traced the detailed 3D dust map published by Green et al. (2015) for the given coordinates, at which distance such a reddening is expected. The reddening is very low until about 300 pc and reaches 0.20 mag at 400 pc and again increases very rapidly at 500 pc.

As next step, we calculated the absolute magnitude using the above listed reddening for a distance between 400 pc and 500 pc. From this, we derive $0.96 < M_V < 0.47$ mag. This is typical for a B9.5 V star at or close to the Zero-Age-Main-Sequence with a mass of about 2.7 M$_\odot$ (Chen et al. 2015). These results are in excellent agreement with those published by Tetzlaff et al. (2011) who listed a mass of $2.9\,M_\odot$ and an age of about 10 Myr.

HD 240121 is listed as a variable in the field of the open cluster NGC 7429 by Zejda et al. (2012). The cluster parameters of NGC 7429 became available after their publication by Kharchenko et al. (2013). They give a distance of 1 250 pc and a reddening of 0.949 mag for NGC 7429. Comparing these values with those for HD 240121, we can definitely exclude a cluster membership.

## 4. Results

HD 240121 has been spectroscopically confirmed as a CP2 star of the silicon subgroup. Its period of light variability ($P = 2.0232$ d), amplitude ($\Delta B = 0.08$ mag; $\Delta V = 0.04$ mag; $\Delta B > \Delta V$) and variability pattern (in particular the anti-phase variations of the $B$ and $V$ light curves) are typical of an ACV variable. From reddening and absolute magnitude estimates, the star's membership to the open cluster NGC 7429 is excluded.

Taking into account all available data, we conclude that HD 240121 is a young, classical, late B-type CP2 star. Such stars are quite rare and excellent astrophysical laboratories to test the formation and evolution of stellar magnetic fields as well as stellar atmospheres. Because of the unusual anti-phase variations of the $B$ and $V$ light curves, which imply a null wavelength in the visual range, multicolour photometric and spectroscopic monitoring of this interesting variable star are strongly encouraged.

## Acknowledgements

This project is financed by the SoMoPro II programme (3SGA5916). The research leading to these results received a financial grant from the People Programme (Marie Curie action) of the Seventh Framework Programme of the EU according to REA Grant Agreement No. 291782. The research is further co-financed by the South-Moravian Region. This work reflects only the authors' views so the European Union is not liable for any use that may be made of the information contained therein. This research has made use of the SIMBAD database and the VizieR catalogue access tool operated at CDS, Strasbourg, France.